\documentclass[apjl]{emulateapj}
\usepackage{apjfonts}
\usepackage{pslatex}
\usepackage{graphicx}

\shorttitle{SZ Effect from Quasars}
\shortauthors{Chatterjee \& Kosowsky}

\journalinfo{Accepted into \apjl~Letters, April 17, 2007}

\begin{document}
\title{The Sunyaev-Zeldovich Effect from Quasar Feedback} 
\author{Suchetana Chatterjee
 \& Arthur Kosowsky
}
\affil{Department of Physics and Astronomy, University of Pittsburgh, Pittsburgh, PA 15260}
\begin{abstract}
The observed relationship between X-ray luminosity and temperature of the diffuse intercluster medium clearly shows the effect of nongravitational heating on the formation of galaxy clusters. Quasar feedback into the intergalactic medium can potentially be an important source of heating, and can have significant impact on structure formation. This feedback process is a source of thermal Sunyaev-Zel'dovich distortions of the cosmic microwave background. Using a simple one-dimensional 
Sedov-Taylor model of energy outflow, we calculate the angular power spectrum of the temperature distortion, which has an amplitude on the order of one micro-Kelvin. This signal will be at the noise limit of upcoming arcminute-scale microwave background experiments, including the Atacama Cosmology Telescope and the South Pole Telescope, but will be directly detectable with deep exposures by the Atacama Large Millimeter Array or by stacking many microwave images.
\end{abstract}

\section{Introduction}
The tiny temperature fluctuations in the cosmic microwave background, arising from small
density perturbations in the early universe, have proven to be our most powerful constraint
on properties of the universe (Spergel et al. 2006), 
spectacularly fulfilling theoretical expectations (Jungman et al.\ 1996a).
The WMAP satellite, the culmination of numerous microwave experiments, has mapped
the microwave sky with an angular resolution of 15 arcminutes (Hinshaw et al. 2006,
Jarosik et al. 2006). 
Attention is now turning to smaller angular scales of a few arcminutes, 
where temperature fluctuations arise
due to interaction of the microwave photons with matter in the low-redshift universe
(for a review, see Kosowsky 2003). These secondary anisotropies arise due to
nonlinear evolution of gravitational potentials (Rees \& Sciama 1968), peculiar velocities
of ionized dense regions (Ostriker \& Vishniac 1986), gravitational lensing (Blanchard \& Schneider 1987, Bernardeau 1997), and microwave scattering from hot electrons (Sunyaev \& Zeldovich 1972). 
The last of these, now known as the Sunyaev-Zeldovich (SZ) Effect, provides a powerful
method for detecting accumulations of hot gas. 

Galaxy clusters, which contain the majority of
the thermal energy in the universe, provide the largest SZ signal, and upcoming
microwave surveys like ACT, the Atacama Cosmology Telescope (Kosowsky et al. 2006), and SPT,
the South Pole Telescope (Ruhl et al. 2004), are expected to detect thousands of clusters
over their entire range of redshift, potentially placing strong constraints
on the evolution of structure in the universe (e.g., White, Efstathiou, \& Frenk 1993, Bahcall, Fan, \& Cen 1997, Molnar et al. 2004, but see also Francis, Bean, and Kosowsky 2005). 
A number of other astrophysical processes will also create measurable SZ distortions,
including peculiar velocities during reionization (McQuinn et al.\ 2005, Iliev et al. 2006),
supernova-driven galactic winds (Majumdar, Nath, \& Chiba 2001), black hole seeded proto-galaxies (Aghanim, Ballad \& Silk 2000), kinetic SZ from Lyman Break Galaxy outflow (Babich \& Loeb 2007), effervescent heating in groups and clusters of galaxies (Roychowdhury, Ruszkowski \& Nath 2005) and supernovae from
the first generation of stars (Oh, Cooray, \& Kamionkowski 2003). Here we investigate one
other generic source for SZ signals: the hot bubble surrounding a quasar or active
galactic nucleus. Previous related studies were done by several authors (e.g., Natarajan \& Sigurdsson 1999, Yamada, Sugiyama \& Silk 1999; Lapi, Cavaliere, \& De Zotti 2003; Platania et al.\ 2002).

Analytic studies and numerical simulations of cluster formation indicate that the temperature and the X-ray luminosity relation should be $L_{x} \simeq T^{2}$ in the absence of gas cooling and heating (Peterson \& Fabian 2006). Observations show instead that $L_{x} \simeq T^{3}$ over the temperature range 2-8 kev with a wide dispersion at lower temperature, and a possible flattening above (Markevitch 1998, Arnaud \& Evrard 1999, Peterson \& Fabian 2006). The simplest explanation for this result is that the gas had an additional heating of 2-3 keV per particle (Wu, Fabian \& Nulsen 2000; Voit et al. 2003). Several nonthermal heating sources have been discussed in this context (Peterson \& Fabian 2006), and quasar outflow (Nath \& Roychowdhury 2002) is one possibility. The effect of this feedback mechanism on different scales of structure formation have been addressed by several authors (e.g., Mo \& Mao 2002; Oh \& Benson 2003; Granato et al. 2003). Recent progress in understanding the relation of black hole masses to dark matter haloes (Merritt \& Ferrarese 2001; Tremaine et al.\ 2002) provides a simple
connection between individual quasars and their overall cosmological impact; Scannapieco \& Oh (2004) have studied the global impact of quasar feedback into the IGM on structure formation in the standard cosmological model. In this paper, we use this model to investigate the quasar feedback process as a source of microwave background SZ distortion, and discuss its detectability by 
upcoming SZ surveys and millimeter wave experiments.

\section{Quasar Outflow Model}

A quasar injects a substantial amount of energy into the surrounding gas while it is
active. Following Scannapieco \& Oh (2004), we assume the black hole powering the quasar
shines at its Eddington luminosity and returns around 5\% of this energy to the galactic gas, eventually disrupting its own fuel source after a dynamical time of the cold gas surrounding the
black hole, $t_{\rm dyn} \simeq 5\times10^{7}(1+z)^{-3/2}$ yr (Barkana \& Loeb 2001).  The efficiency
factor is estimated by requiring that the energy is sufficient to explain the preheating inferred from galaxy cluster X-ray observations. The duration of the energy injection 
$t_{\rm dyn}$ is much shorter than the expansion time of the resulting bubble of hot gas (on the order of 
$10^{9}$ years), so we assume an instantaneous point source injection of energy into the
intergalactic medium.  
The total energy input is just the product of the luminosity, the efficiency factor, and
the duration. The dependence of the luminosity on black hole mass can be converted to
a dependency on galaxy halo mass via the $M_{bh}$-$\sigma$ relation (Merritt \& Ferrarese 2001; Tremaine et al.\ 2002) and a relation between galaxy rotational velocity and velocity
dispersion (Ferrarese 2002). The resulting total energy injected for a quasar turning on
at redshift $z$ in a halo of mass $M_{\rm halo}$ is approximately
$E = 0.06M_{12}^{5/3} (1+z)$,
where $M_{12}\equiv M_{\rm halo}/10^{12}M_\odot$. 

For simplicity, we assume that after the energy injection, a hot bubble evolves adiabatically and expands into a medium of uniform overdensity. 
The one-dimensional Sedov-Taylor solution can be used to model the radius and temperature of the region contained by the blast wave (Scannapieco \& Oh 2004). 
The radius scales as 
\begin{equation} 
R_{s} = 1.7E_{60}^{1/5}\delta_{s}^{-1/5}(1+z)^{-3/5} t_{\rm Gyr}^{2/5} \,\, \rm{Mpc},
\end{equation} 
where $E_{60}$ is the energy in the hot medium $E$ in units of $10^{60}$ ergs, $\delta_{s}$ is the ratio of the density of the surrounding medium to the mean cosmic baryon density, and $t_{\rm Gyr}$ is the expansion time of the bubble in units of $10^9$ years. We have assumed $\delta_{s}$ equal to one which means, an ambient baryon density equal to the cosmological density. Many quasars may reside in denser environments, but the y-distortion scales only very weakly with density. We also have assumed an exact proportionality between halo and black hole mass; some quasars may have substantially higher energy outputs. The temperature of the bubble scales as 
\begin{equation} 
T_{s} = 3.1\times10^{7}E_{60}\delta_{s}^{-1}(1+z)^{-3}\left(\frac{R_s}{1\,{\rm Mpc}}\right)^{-3} \,\, \rm{K}.
\end{equation} 

The density of the gas inside the bubble is assumed to be uniform and equal to the density of the gas outside the bubble. The actual density profile varies with radius (e.g., Shu 1992), but not strongly, and for simplicity we assumed it to be constant.
The cooling time of the gas is of the order of a Hubble time, so we ignore radiative cooling (cf.\ Figure 2 of Scannapieco \& Oh 2004). Energy loss into  compression against gas
pressure and gravitational potential energy changes are also ignored. We also assume that all quasars eject their energy at a single redshift, $z_{in}$. A more realistic model would integrate over
a range of input redshifts, but this simple assumption will give the correct order of magnitude for
the resulting SZ distortions with this mean redshift, since the SZ effect is essentially independent of redshift. 

\section{The Sunyaev-Zeldovich Distortion}

The Compton $y$-parameter characterizing the SZ spectral distortion is proportional to
the line-of-sight integral of the electron pressure. 
Since each individual source is assumed to be spherically symmetric, with constant
temperature $T_e$ and number density $n_e$ of electrons inside the hot bubble surrounding the quasar,
the y-distortion $y(\theta)$ on the sky will be azimuthally symmetric, depending only on the
angle $\theta$ between the bubble center and a particular line of sight. 
The angular Fourier transform of the y-distortion is given by (cf.\ Peebles 1980)
\begin{equation} 
y_{l} = \frac{8\pi\sigma_T}{m_e} T_e n_e R_s \int \theta d \theta\left[1-\frac{D_A^2\theta^2}{R_s^2}\right]^{1/2}
J_{0}\left[\left(l+\frac{1}{2}\right)\theta\right],
\end{equation} 
with $\sigma_T$ the Thompson cross section and $D_A(z)$ the angular diameter distance
to redshift $z$. The square-root term in the integrand is the path length
through the hot bubble at an angle $\theta$ from its center. This integral
can be performed analytically (Gradshteyn \& Ryzhik 1980) to give
\begin{equation}
y_l(M,z) = \frac{16\sigma_T T_e n_e R_s^{3/2}}{D_A^{1/2}} \left(\frac{\pi}{2l+1}\right)^{3/2}
J_{3/2}\left[\left(l+\frac{1}{2}\right)\frac{R_s}{D_A}\right].
\end{equation}
Note that $T_e$ and $R_s$ depend on both the halo mass $M$ and the redshift $z$, and
$n_e$ depends on $z$. 

The y-distortion on the sky can be conventionally expanded in terms of the spherical harmonics as 
$y(\hat{\bf n}) = \sum_{l m}a_{l m}Y_{lm}(\hat{\bf n})$. The angular power spectrum is then obtained as
$C_{l} = \langle\left|a_{lm}\right|^{2}\rangle$, an ensemble average over the coefficients. The power spectrum has two components (Cole \& Kaiser 1988, Cooray \& Sheth 2002), $C^{yy}_{l} = C_{l}^{p} + C_{l}^{c}$, where 
$C_{l}^{p}$ is the contribution from Poisson noise of the random galaxy distribution, and $C_{l}^{c}$ comes from the correlation between galaxies. The two terms are given as (Komatsu \& Kityama 1999; Majumder, Nath \& Chiba 2000)
\begin{equation} 
C_{l}^p =\int_{0}^{z_{in}}dz\frac{dV}{dz}\int_{M_{min}}^{M_{max}}dM\frac{dn(M,z_{in})}{dM}\left| y_{l}(M,z)\right|^{2},
\end{equation} 
\begin{equation} C_{l}^c =\int_{0}^{z_{in}}dz\frac{dV}{dz}P_{m}(k_l(z))\left(\int_{M_{min}}^{M_{max}}dM\,\Phi_{l}(M,z)\right)^{2}
\end{equation} 
where 
\begin{equation} 
\Phi_{l}(M,z)=\frac{dn(M,z_{in})}{dM}b(M,z_{in})y_{l}(M,z),
\end{equation} 
$k_l(z)\equiv l/D_A(z)$ is the wavenumber corresponding to the multipole angular scale $l$ at
redshift $z$,
$dV/dz$ is the comoving volume element, $dn(M,z)/dM$ is the differential mass function, 
$P_{m}(k,z)$ is the matter power spectrum, and $b(M,z)$ is the linear bias factor. The expression
for the correlated piece uses the Limber approximation. 

To determine the cosmological functions, we assume a 
standard $\Lambda$CDM cosmology with $\Omega_{m}=0.31$, $\Omega_{b}=0.044$, and a
Harrison-Zeldovich primordial power spectrum $P(k)=k$. The matter power spectrum is computed using the transfer function fits given by Eisenstein \& Hu (1999). The power spectrum is normalized to 
the WMAP3 value $\sigma_{8} = 0.77$ (Spergel et al.\ 2006). 

\section{Power Spectra and Mean Signals}

Figure 1 shows the y-distortion power spectrum, for different values of the energy
injection redshift $z_{in}$. The correlated term
dominates for $l\lesssim 10^4$, with a broad, relatively flat contribution between $l=100$
and $l=2000$, corresponding to angular scales from 2 degrees down to 5 arcminutes (the
angular scales on which large scale structure is evident). The
Poisson term contributes the secondary peak around $l=3\times 10^4$, at an angular
scale of around $20''$ (the characteristic separation of galaxies). Reducing
the energy input redshift $z_{in}$ from 3 to 2.5 reduces the power spectrum by roughly
a factor of 2, with the Poisson term and the correlated contribution being affected
about equally. The dependence on maximum mass $M_{\rm max}$ is realtively weak: the power
spectrum amplitude increases only by a factor of around 60\% if the maximum
mass is increased by a factor of 5. For comparison, Fig.~1 also shows the primary
microwave background anisotropy, and the noise per $l$ value for a model ACT-like
experiment which maps 400 square degrees at one arcminute resolution and noise
of 2 $\mu$K per pixel, using the approximate formula in Jungman et al.\ 1996b. Note the
signal discussed here is above the noise level for a range in $l$. Other signals will
also contribute, particularly galaxy cluster SZ distortions (Komatsu and Seljak 2002). 

\begin{figure}
\includegraphics[width=8cm]{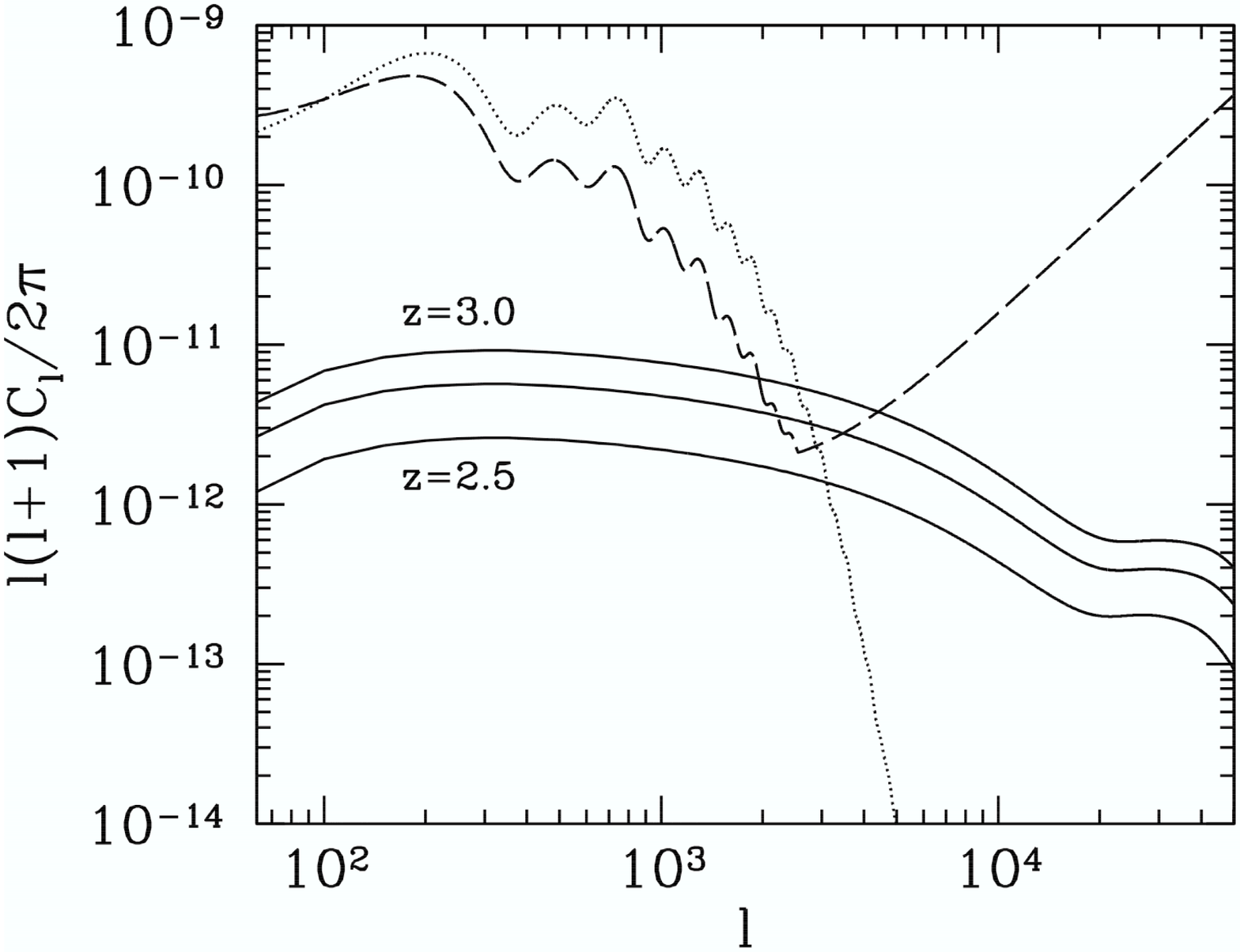}
\caption{The y-distortion power spectrum, for $M_{\rm max}=2\times 10^{12} M_\odot$ and 
  $M_{\rm min} =10^{11}M_{\odot}$. The three solid lines from bottom to top are for $z_{in}=2.5$, 
  $2.8$, and $3$. The signal is a combination of the correlation term (which peaks at $l=290$)
  and the Poisson term (peaking at $l=3.6\times 10^4$). Also plotted is the primary anisotropy
  (dotted line) and the noise level per $l$ value (dashed line) for an ACT-like model experiment covering
  400 square degrees with 1 arcminute resolution and a pixel noise of 2 $\mu$K.}
\label{fig:z_in}
\end{figure}

Root-mean-square temperature fluctuations are obtained by converting from y-distortion to temperature 
power spectrum at a given frequency, and then convolving with a gaussian beam profile. We consider three different beam widths: an ACT-like beam of 1 arcminute,
and 2 ALMA resolutions of 15 and 5 arcseconds.  The results are shown in Table 1, for
the power spectrum with $z_{in}=3$, $M_{\rm max}=2\times 10^{12} M_\odot$, and
$M_{\rm min}=10^{11} M_\odot$, corresponding to the solid line in Fig.~1. Nominal
noise targets for ACT are a few $\mu$K per arcminute pixel (Kosowsky et al. 2006); 
the signal calculated here will be 
at the detector noise threshold. For ALMA, $1\sigma$ sensitivities in continuum brightness
are given in Table 2, for a 50-antenna configuration, 
an integration time of 1 hour and a velocity resolution of 1000 km/s. These
are calculated using the sensitivity formulae given in ALMA memo 393 (Guilloteau 2003).
With an integration of this length, the SZ signal from quasars is also near the noise
limit for a compact configuration with 15 arcsecond resolution; substantially
longer integrations will achieve raw sensitivity for detection of the signal.
\begin{table}
\begin{center}
\begin{tabular}[t]{c|c|c}
\hline
\multicolumn{1}{c|}{Frequency}&
\multicolumn{1}{c|}{Resolution}&
\multicolumn{1}{c}{Temperature}\\
\multicolumn{1}{c|}{(GHz)}&
\multicolumn{1}{c|}{(arcseconds)}&
\multicolumn{1}{c}{($\mu$K)}\\
\hline
 145 & 60 & 2.18 \\ 
 220 & 60 & 0.09 \\
 265 & 60 & 1.63 \\ 
 145 & 15 & 2.32\\ 
 220 & 15 & 0.11\\ 
 265 & 15 & 1.75\\
 145 & 5 & 2.35\\
 220 & 5 & 0.11\\ 
 265 & 5 & 1.78\\\hline 
\end{tabular}
\end{center}
\caption{Root-mean-square temperature fluctuations at ACT frequencies
and three angular resolutions.}
\label{mean_signals}
\end{table}

\begin{table}
\begin{center}
\begin{tabular}[t]{c|c|c|c}
\hline
\multicolumn{1}{c|}{Frequency}&
\multicolumn{1}{c|}{Resolution}&
\multicolumn{1}{c|}{Baseline}&
\multicolumn{1}{c}{Sensitivity}\\
\multicolumn{1}{c|}{(GHz)}&
\multicolumn{1}{c|}{(arcseconds)}&
\multicolumn{1}{c|}{(km)}&
\multicolumn{1}{c}{($\mu$K)}\\
\hline
 145 & 15 & 0.0284 & 2.41 \\ 
 145 & 5 & 0.0853 & 21.74 \\
 220 & 15 & 0.0187 & 1.76 \\ 
 220 & 5 & 0.0562 & 15.84 \\ 
 265 & 15 & 0.0156 & 1.63 \\
 265 & 5 & 0.0467 & 14.63\\ \hline
\end{tabular} 
\end{center}
\caption{ALMA continuum brightness sensitivities for a one-hour
observation.}
\label{alma_sens}
\end{table}

\section{Discussion}

Quasar feedback is an important issue for cosmological evolution because of its potential for providing nongravitational heating and its global impact on structure formation. The resulting Sunyaev-Zeldovich distortion of the microwave background due to this feedback process is a robust observational consequence of this theoretical scenario. Similar signals due
to supernovas from the first generation of stars have previously been investigated
(Oh, Cooray, \& Kamionkowski 2003), with an amplitude the same order of magnitude
as in Fig.~1.  Both of these signals, arising from the superposition of many
small-angle SZ distortions, result in a small net amplitude temperature fluctuation spread broadly
in angular power. The signal is quite interesting in its own right, and also gives a possible
noise source for current arcminute-resolution microwave experiments.

 APEX and SPT both aim to produce maps of the microwave sky with detector
noise of around 10 $\mu$K per arcminute sky pixel. The SZ distortions from quasars 
is well below this noise level, and should not be an important contributor to the noise
budget for these experiments. ACT initially plans to cover less sky at greater depth,
with a pixel noise of a few $\mu$K. If ACT 
eventually attains a nominal map sensitivity of 2 $\mu$K per pixel, the quasar
SZ signal will be at the noise limit. Also note that
infrared emission from high-redshift dusty galaxies, as detected by SCUBA (Borys et al. 2003),
will provide a confusion noise limit somewhere between 1 and 10 $\mu$K at
ACT and SPT frequencies and resolutions. While individual quasars will not be detected by these experiments, stacking
microwave images centered on known quasars can give a high significance detection.
For a map with noise of 10 $\mu$K per 1' beam, each quasar SZ signal will have
a signal-to-noise ratio of around 0.2. Data from the Sloan Digital Sky Survey
reveal around 50 photometrically detected quasars per square degree (Richards et al.\ 2006);
a 200 square-degree microwave survey with optical imaging will give around $10^4$
quasars. Stacking map sections centered on $N$ quasars will increase the signal-noise
ratio by a factor of $N^{1/2}$, giving a 20$\sigma$ detection of the quasar SZ distortions. 
The actual sky area needed could be much smaller, since many low-redshift galaxies once hosted
quasars, and the cooling time of the hot bubbles surrounding them is long. 
Distinguishing this signal from any intrinsic emission may require measurements at multiple
frequencies or increased angular resolution to resolve the sources.

Interferometric experiments like ALMA will have superior spatial resolution and, due to
a large number of telescopes, very high potential sensitivity. ALMA's spatial resolution
will be sufficient to resolve the intergalactic separation and thus see individual galaxies,
reducing any confusion noise. 
In a high-resolution configuration, ALMA could distinguish between the SZ signal from a hot
bubble surrounding a galaxy and infrared emission coming directly from dust in the galaxy
or from a dusty torus surrounding a quasar (Kawakatu et al., 2007). 
The ALMA design reference science plan (Dishoeck, Wootten, \& Guilloteau 2003) proposes 
over one year of total observations devoted to galaxies and cosmology, likely including
a deep survey in several bands which would be useful for detecting the signal
considered here. The angular resolutions of 5 and 15 arcseconds in Table 2 are a good match for the angular size of the hot bubbles, which is around 10 arcseconds. Substantially larger beam sizes will not result in increased sensitivity, since more noise will be added from the regions outside of the unresolved sources, while smaller beam sizes will give a lower raw sensitivity for a given integration time. 

The prospect of clear detection of SZ distortions from quasar energy output is exciting. 
Such observations will probe models of heating of the intergalactic medium, constrain the
thermal history of the universe at low redshifts, and provide important inputs into models
of galaxy and quasar evolution. We hope these results will prompt more detailed investigation of
both the signal and its detectability with ALMA and upcoming microwave experiments.

\acknowledgments
We thank Evan Scannapieco, Eichiro Komatsu, Ryan Scranton and Tim Hamilton for helpful discussions, and Sandhya Rao and Dave Turnshek for careful readings of the manuscript.
Andrew Blain provided expert guidance about ALMA and its capabilities. Special thanks to
James Aguirre, David Spergel, and Shirley Ho who suggested stacking images to detect the
signal in microwave maps and the referee for some other useful suggestions. We acknowledge Antony Lewis and Anthony Challinor for CAMB, which has been used to obtain the primary anisotropy. This work was supported by the National Science Foundation through grant AST-0408698 to the ACT project, and by grant AST-0546035.

\end{document}